\documentclass[10pt,pra,aps,twocolumn,showpacs,amsmath,amssymb,superscriptaddress]{revtex4-1}

\usepackage{times}
\usepackage{bm}
\usepackage{textcomp}
\usepackage{graphicx}
\usepackage{bm}
\allowdisplaybreaks 
\usepackage{braket}
\usepackage{physics}
\usepackage{subdepth}
\usepackage[breaklinks]{hyperref}
\usepackage{todonotes}
\newcommand{\pd}{\phantom{\dag}}

\hypersetup{colorlinks=true, linkcolor=blue, citecolor=blue, filecolor=blue, urlcolor=blue}

\newcommand{\RN}[1]{\textup{\uppercase\expandafter{\romannumeral#1}}}

\begin{document}

\title{Resonant inelastic X-ray scattering in topological semimetal FeSi}

\author{Yao Shen}

\affiliation{Condensed Matter Physics and Materials Science Department, Brookhaven National Laboratory, Upton, New York 11973, USA}

\author{Anirudh Chandrasekaran}

\affiliation{Department of Physics, Boston University, Boston, MA, 02215, USA}
\affiliation{Department of Physics and Centre for the Science of Materials, Loughborough University, Loughborough LE11 3TU, UK}

\author{Jennifer Sears}
\affiliation{Condensed Matter Physics and Materials Science Department, Brookhaven National Laboratory, Upton, New York 11973, USA}

\author{Tiantian Zhang}
\affiliation{Department of Physics, Tokyo Institute of Technology, Okayama, Meguro-ku, Tokyo, Japan}  % she did the DFT
\affiliation{Tokodai Institute for Element Strategy, Tokyo Institute of Technology, Nagatsuta, Midori-ku, Yokohama, Kanagawa, Japan}

\author{Xin Han}
\author{Youguo Shi}
\affiliation{Beijing National Laboratory for Condensed Matter Physics, Institute of Physics, Chinese Academy of Sciences, Beijing 100190, China}

\author{Jiemin Li}
\author{Jonathan Pelliciari}
\author{Valentina Bisogni}
\affiliation{National Synchrotron Light Source II, Brookhaven National Laboratory, Upton, New York 11973, USA}

\author{Mark P.~M.~Dean}\email[]{mdean@bnl.gov}

\affiliation{Condensed Matter Physics and Materials Science Department, Brookhaven National Laboratory, Upton, New York 11973, USA}

\author{Stefanos Kourtis}\email[]{Stefanos.Kourtis@usherbrooke.ca}

\affiliation{Institut quantique \& D\'{e}partement de physique, Universit\'{e} de Sherbrooke, J1K 2R1, Qu\'{e}bec, Canada}
\affiliation{Department of Physics, Boston University, Boston, MA, 02215, USA}

\begin{abstract}
The energy spectrum of topological semimetals contains protected degeneracies in reciprocal space that correspond to Weyl, Dirac, or multifold fermionic states. To exploit the unconventional properties of these states, one has to access the electronic structure of the three-dimensional bulk. In this work, we present the first joint theory-experiment study of the electronic structure of a candidate topological semimetal with resonant inelastic X-ray scattering (RIXS). We resolve the bulk electronic states of FeSi using momentum-dependent RIXS at the Fe $L_3$ edge. We observe a broad excitation continuum devoid of sharp features, consistent with particle-hole scattering in an underlying electronic band structure. Using density functional theory (DFT), we calculate the electronic structure of FeSi and derive a band theory formulation of RIXS in the fast collision approximation to model the scattering process with zero adjustable parameters. While band theory predicts an excitation continuum with broad spectral features similar to the observed ones, discrepancies between theory and experiment suggest the presence of low-energy processes that DFT alone does not account for. This first study of RIXS in a topological semimetal shows that RIXS is a useful tool for revealing unanticipated behavior of bulk electronic states in this class of materials.
\end{abstract}

\date{\rm\today}

\maketitle

\section{Introduction}

In the last decade and a half, topological matter has become a cornerstone of quantum materials science~\cite{hasan2010colloquium}. The discovery of three-dimensional topological insulators~\cite{hasan2011three}, in particular, sparked a flurry of activity in the then nascent field. Electrons in these crystalline materials are effectively noninteracting, giving rise to electronic bands in the bulk that are indistinguishable from those of a trivial band insulator. The electronic wavefunction, however, is characterized by topological indices that dictate the presence of symmetry-protected Dirac states at the surface of the material, as well as nontrivial (magneto-)transport responses.

More recently, topological semimetals have been added to the catalogue of three-dimensional topological materials~\cite{burkov2016topological,armitage2018}. These systems also feature topologically protected boundary states and nontrivial (magneto-)transport, but additionally have distinct geometric characteristics in their bulk band structure. In the simplest case of Weyl semimetals, these geometric characteristics are singly degenerate energy surfaces in reciprocal space that contain a band touching point---a Weyl node---around which electronic bands disperse linearly in all three directions in reciprocal space~\cite{VonNeumann1993,Herring1937,Blount1962}. Such band touchings are Berry curvature singularities characterized by topological indices. The value of the topological index of a nodal point determines the geometry of the band dispersion in the vicinity of nodal points~\cite{Onoda2002,Fang2012,bradlyn2016beyond}. Conversely, the geometry of the bulk bands becomes a proxy of topology in these materials. Measuring the electronic density of states in the bulk can therefore reveal the topological nature of a semimetal.

Resonant inelastic X-ray scattering (RIXS) is a spectroscopic technique that yields momentum- and energy-resolved spectra of charge-neutral electronic excitations. While RIXS has been extensively used in studying magnetic excitations in gapped materials like insulators and superconductors~\cite{fatale2015magnetic, Dean2015insights, Mitrano2024exploring}, it is increasingly applied in studies of compounds that host itinerant carriers with small or no charge gaps~\cite{Brookes2020spin, Pelliciari2021tuning, jia2022interplay}. That RIXS can be used to map electronic bands of materials, including semimetals, has long been established~\cite{Ma1994,Johnson1994a,carlisle1995,carlisle1999,denlinger2002,kokko2003,strocov2004a,zhang2012electronic,monney2020mapping}. Improvements in resolution in recent years have renewed interest in using RIXS to detect band structure effects at meV energy scales in materials of technological interest, such as unconventional superconductors~\cite{kanasz2016resonant}. There are even theoretical proposals to use RIXS to measure topological indices of nodal points in topological semimetals~\cite{kourtis2016a,schueler2022probing}. These prospects are particularly appealing for probing the bulk of three-dimensional materials, since alternative techniques such as angle-resolved photoemission (ARPES) and scanning tunneling spectroscopy, probe predominantly the surface rather than the bulk. Furthermore, topological nodal points may only appear above the Fermi level or as a result of an applied magnetic field, settings in which the resolving power of ARPES is limited. As these settings may be relevant to the technological exploitation of topological materials, alternative methods to visualize the bulk band structure and to identify topological features are sought after. Before honing in on properties of topological origin, however, one has to determine whether bulk band structure effects at large are detectable in RIXS of topological semimetals.

The monosilicide family $X$Si with $X = $ Fe, Co, Rh, Mn, Re, Ru was proposed as a platform for the realization of multifold fermions, a generalization of Weyl fermion states arising at points of higher degeneracy in a band structure~\cite{bradlyn2016beyond,tang2017multiple}. These materials feature the cubic B20-type structure with noncentrosymmetric space group $P2_13$ (no.~198). While some studies report success in modeling FeSi as a nonmagnetic band insulator or semiconductor with a gap of 50-80 meV solely with DFT~\cite{Changdar2020electronic, Ohtsuka2021emergence}, others report significant band renormalization effects due to interactions~\cite{arita2008angle}. Furthermore, many experimental studies of FeSi have revealed an intricate temperature dependence of electronic properties~\cite{Changdar2020electronic, petrova2010elastic, arita2008angle, schlesinger1993unconventional}. Finally, it has recently been debated whether FeSi exhibits Fermi arcs near the Fermi level, even within the band structure picture~\cite{Changdar2020electronic}. These intricacies of the electronic properties of FeSi warrant additional research on this material with different methods.

In this work, we use RIXS to probe the bulk of FeSi, aiming to quantitatively test how bulk band structure manifests in RIXS spectra of a putative topological semimetal. We observe broad continua in the RIXS spectra of FeSi, consistent with particle-hole scattering in an underlying band structure. We model the RIXS process in the fast-collision approximation using the band structure of FeSi as determined by density functional theory (DFT) calculations. Our modeling is completely \emph{ab initio} and includes zero adjustable parameters, yet captures the number and position of the most prominent RIXS spectral features, albeit imperfectly. This agreement is comparable to the state of the art in spectroscopy of topological semimetals. Our results show that discrepancies between experimental RIXS spectra and band theory of candidate topological semimetals may be significant. Understanding the origin of these discrepancies is an important stepping stone towards higher resolution RIXS experiments aiming to visualize topological nodal points, and thus identify and classify topological semimetals.

\section{Theory}\label{sec:rixs-theory}

\subsection{RIXS cross section and fast collision approximation}

\begin{figure*}[t]
 \centering
 \includegraphics[width=0.8\textwidth]{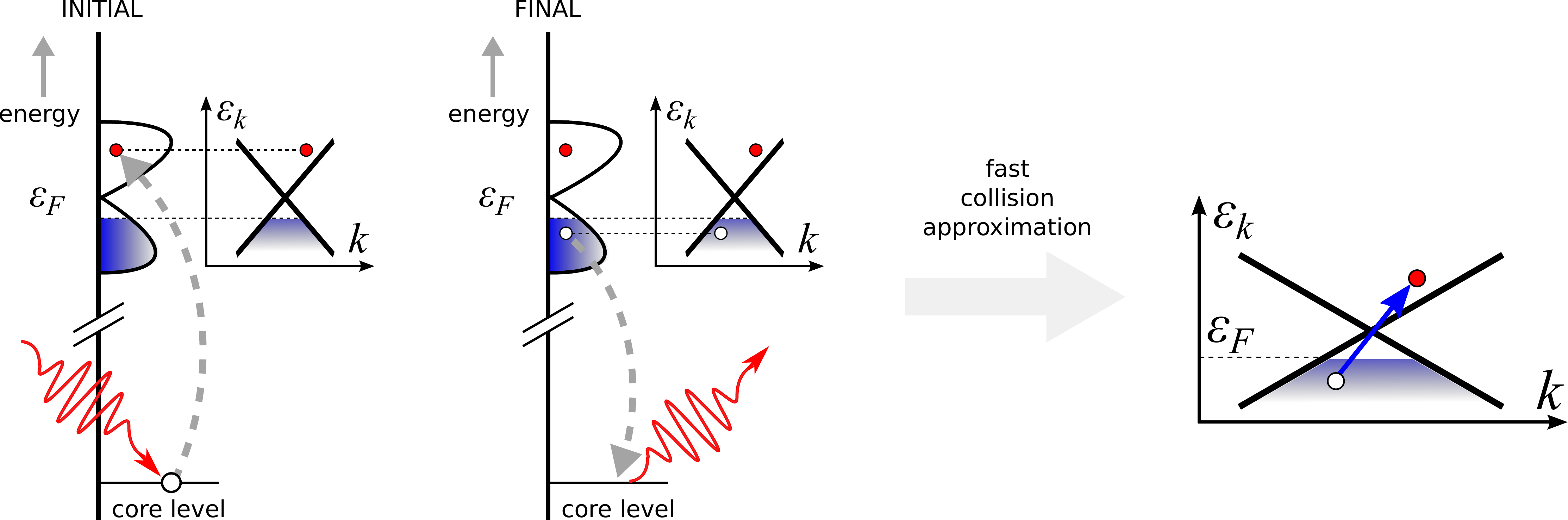}
 \caption{Illustration of the direct RIXS process and reduction to effective particle-hole scattering via the fast-collision approximation. } 
 \label{fig:cartoon}
\end{figure*}

We briefly introduce the theoretical description of RIXS at zero temperature. More comprehensive presentations of RIXS can be found in Refs.~\onlinecite{Kotani2001,Ament2011}.

In a RIXS experiment, core electrons of an ion are promoted to a state above the Fermi level $\varepsilon_F$ by an intense x-ray beam, thereby locally exciting the irradiated material into a highly energetic and short-lived intermediate state. Subsequently, the core hole recombines with a valence electron. The process imparts both energy and momentum to particle-hole excitations in the material. In what follows, we will consider excitation of core electrons directly into orbital(s) close to $\varepsilon_F$, which give rise to the low-energy physics in the material. This process, which is often referred to as \textit{direct} RIXS, is illustrated in Fig.~\ref{fig:cartoon}.

The double differential cross section is a measure of the total RIXS intensity. Up to a constant prefactor, it is given by
\begin{multline}
 I(\vb{k}_{\text{in}}^{\,},\vb{k}_{\text{out}}^{\,},\omega_{\text{in}}^{\,},\omega_{\text{out}}^{\,},\boldsymbol{\epsilon}_{\text{in}}^{\,},\boldsymbol{\epsilon}_{\text{out}}^{\,}) \\
  = \sum_{fg} |{\cal F}_{fg}(\vb{k}_{\text{in}}^{\,},\vb{k}_{\text{out}}^{\,},\omega_{\text{in}}^{\,},\boldsymbol{\epsilon}_{\text{in}}^{\,},\boldsymbol{\epsilon}_{\text{out}}^{\,})|^2 \delta(E_g - E_f + \hbar \Delta \omega) \,,\label{eq:cross}
\end{multline}
where $\hbar \Delta \omega = \hbar(\omega_{\text{in}}^{\,} - \omega_{\text{out}}^{\,})$ is the energy transferred to the material, $\vb{k}_{\text{in}}^{\,}$ and $\vb{k}_{\text{out}}^{\,}$ ($\boldsymbol\epsilon_{\text{in}}^{\,}$ and $\boldsymbol\epsilon_{\text{out}}^{\,}$) the incoming and outgoing photon wavevectors (polarizations) and $E_g$ and $E_f$ the energies corresponding to initial and final many-body states $\ket{g}$ and $\ket{f}$ of the valence electrons. The scattering amplitude ${\cal F}_{fg}$ in the dipole approximation is
\begin{multline}
 {\cal F}_{fg}(\vb{k}_{\text{in}}^{\,},\vb{k}_{\text{out}}^{\,},\omega_{\text{in}}^{\,},\boldsymbol\epsilon_{\text{in}}^{\,},\boldsymbol\epsilon_{\text{out}}^{\,}) \\ = \langle f| \widehat{\cal D}^{\dag}(\boldsymbol\epsilon_{\text{out}}^{\,},\vb{k}_{\text{out}}^{\,}) \, \widehat{\cal G}(\omega_{\text{in}}^{\,}) \, \widehat{\cal D}(\boldsymbol\epsilon_{\text{in}}^{\,},\vb{k}_{\text{in}}^{\,}) |g \rangle \,, \label{eq:scamp}
\end{multline}
where $\widehat{\cal G}$ is the intermediate-state propagator
\begin{equation}
 \widehat{\cal G}(\omega_{\text{in}}^{\,}) = (E_g + \hbar\omega_{\text{in}}^{\,} + \text{i}\Gamma - \widehat{\cal H})^{-1} \,,
\end{equation}
with $\widehat{\cal H}$ the Hamiltonian describing the system in the intermediate excited state and $\Gamma$ the intermediate-state inverse lifetime. The dipole operators $\widehat{\cal D}$ and $\widehat{\cal D}^{\, \dag}$ represent the x-ray absorption and emission, respectively. For a crystalline material, they can be written as
\begin{align}
 \widehat{\cal D}(\boldsymbol\epsilon,\vb{k}) =&{\ } \boldsymbol{\epsilon \cdot} \widehat{\vb{D}}_{\vb{k}} \,,\\
 \widehat{\vb{D}}_{\vb{k}} = &{\ } \sum_{\mu,\nu} \langle \mu| \widehat{\vb{r}} - \vb{r}_{\mu} |\nu \rangle  \sum_{\vb{R}} e^{\text{i} \vb{k}\cdot\vb{R}} \, \widehat{d}_{\vb{R}\mu}^{\,\, \dag} \, \widehat{p}_{\vb{R}\nu}^{\pd} \,,
\end{align}
where $\vb{R}$ is the lattice position. States $|\mu \rangle$ and $|\nu \rangle$ express single-electron valence and core states respectively. The combined valence (core) index $\mu$ ($\nu$) encodes spin, orbital, and sublattice degrees of freedom. Core states $|\nu \rangle$ are conveniently expressed as atomic orbitals, whereas valence states $|\mu \rangle$ can be appropriately chosen Wannier functions, both localized in space around the same position $\vb{r}_{\mu}$ of each atomic site within the unit cell. The position operator $\widehat{\vb{r}}$, defined with respect to each lattice position $\vb{R}$, is the same for all ions. For the $L_{2/3}$ and $M_{2/3}$ resonant edges, the operators $\widehat{d}_{\vb{R}\mu}^{\, \dag}$ and $\widehat{p}_{\vb{R}\nu}^{\ }$ create a $d$-orbital electron and a $p$-orbital core hole, respectively.

Physical arguments allow us to simplify the RIXS scattering amplitude. First, core holes do not hop appreciably; they are created and annihilated at the same site. Taking this into account, ${\cal F}_{fg}$ becomes~\cite{Ament2011}
\begin{multline}
 {\cal F}_{fg}(\vb{q},\omega_{\text{in}}^{\,},\boldsymbol\epsilon_{\text{in}}^{\,},\boldsymbol\epsilon_{\text{out}}^{\,}) \\ = \sum_{\mu,\nu,\mu',\nu'} T_{\mu \nu \mu' \nu'}(\boldsymbol\epsilon_{\text{in}}^{\,},\boldsymbol\epsilon_{\text{out}}^{\,}) F_{\mu \nu \mu' \nu'}(\vb{q},\omega_{\text{in}}^{\,}) \,, \label{eq:split}
\end{multline}
where $\vb{q} = \vb{k}_{\text{in}}^{\,} - \vb{k}_{\text{out}}^{\,}$. The scattering amplitude has been factored in the \textit{atomic scattering tensor}
\begin{equation}
 T_{\mu \nu \mu' \nu'}(\boldsymbol\epsilon_{\text{in}}^{\,},\boldsymbol\epsilon_{\text{out}}^{\,}) = \langle \mu| \boldsymbol\epsilon_{\text{out}}^{\,} \cdot \widehat{\vb{r}} |\nu \rangle^* \langle \mu'| \boldsymbol\epsilon_{\text{in}}^{\,} \cdot \widehat{\vb{r}} |\nu' \rangle \label{eq:geom}
\end{equation}
and the \textit{fundamental scattering amplitude}
\begin{multline}
 F_{\mu \nu \mu' \nu'}(\vb{q},\omega_{\text{in}}^{\,}) \\ = \langle f| \sum_{\vb{R}} e^{-\text{i} \vb{q}\cdot\vb{R}} \, \widehat{p}_{\vb{R}\nu}^{\, \dag} \, \widehat{d}_{\vb{R}\mu}^{\pd} \, \widehat{\cal G}(\omega_{\text{in}}^{\,}) \, \widehat{d}_{\vb{R}\mu'}^{\,\, \dag} \, \widehat{p}_{\vb{R}\nu'}^{\pd} |g \rangle \,. \label{eq:fundamental}
\end{multline}
The intrinsic spectral characteristics of a material are carried by the tensor $F$, which is typically the main quantity of interest in theoretical studies. The tensor $T$ modulates the scattering amplitude according to the geometry of the localized core and valence states. The entries of $T$ can be calculated given knowledge of the valency of the targeted ion and the symmetry group of the crystal~\cite{Haverkort2010,Haverkort2010b,Ament2010c,Kim2017b}. 

Then, within the \emph{fast-collision approximation}, one assumes that $\widehat{\cal G}(\omega_{\text{in}}^{\,}) \approx 1/\Gamma$, where $\Gamma$ is the inverse core-hole lifetime. In this approximation, the RIXS process reduces to the introduction of a particle-hole excitation with fixed momentum and energy in the material --- see Fig~\ref{fig:cartoon} for an example.

Before proceeding to derive the theory of RIXS in band structures, we evaluate the geometric modulation of the RIXS spectrum owing purely to the orbital content of the quantum states involved in the RIXS process. This is obtained by setting the fundamental scattering amplitude $F_{\mu \nu \mu' \nu'}$ to unity:
\begin{equation}
\mathcal{T}(\boldsymbol\epsilon_{\text{in}}^{\,},\boldsymbol\epsilon_{\text{out}}^{\,}) = \left| \sum_{\mu, \nu, \mu', \nu'} T_{\mu \nu \mu' \nu'} (\boldsymbol\epsilon_{\text{in}}^{\,},\boldsymbol\epsilon_{\text{out}}^{\,}) \right|^2 .
\label{eq:rixs_intensity_orbital}
\end{equation}
We shall use $\mathcal{T}$ as a diagnostic to disentangle contributions to the modulation of the RIXS intensity as a function of scattering angles. The first contribution comes through the polarization vectors, which are angle-dependent --- see Fig.~\ref{fig:exp_scheme}. The second contribution is the intrinsic momentum dependence coming from electronic dispersion in the material. In Sec.~\ref{sec:results} we calculate the geometric modulation in Eq.~\eqref{eq:rixs_intensity_orbital} for FeSi and compare it to the scattering angle dependence of RIXS intensity.

\subsection{RIXS process in a band structure}\label{sec:theory}

We wish to describe the RIXS response of crystalline materials in which electrons are, to a good approximation, noninteracting. Valence electrons in these materials are well-described by band theory. The states $\ket{g}$ and $\ket{f}$ in Eq.~\eqref{eq:scamp} are then collections of Bloch modes. 

For a given RIXS edge, one then sums over core states $\ket{\nu}$ and valence Wannier states $\ket{\mu}$ and $\ket{\mu'}$ connected by the dipole operators $\mathcal{\widehat{D}}, \mathcal{\widehat{D}}^\dagger$. Here we study the Fe $L_3$ edge, hence we consider $2p_{3/2}$ orbitals for core electrons and the $3d$ shell for valence electrons. 

In $\vb{k}$-space, the band eigenbasis is given by a unitary rotation of a basis of Wannier states $|\mu \rangle$ per lattice position $\textbf{R}$ to a basis of Bloch states $|\vb{k} \mu \rangle$. The Wannier states have wavefunctions $\varphi_{\mu}^{\phantom \dag} (\vb{x}) = \langle \vb{x} | \mu \rangle$ that are centered about different points in the unit cell, possibly atomic sites. Let $\varphi_{\vb{k}\mu}^{\phantom \dag}(\vb{x}) = \langle \vb{x} |\vb{k}\mu \rangle$ be the spatial wavefunction of $|\vb{k}\mu \rangle$, which could be a spinor. We then have
\begin{equation}
\varphi_{\vb{k}\mu}^{\phantom \dag} (\vb{x}) = \frac{1}{\sqrt{N}} \sum_{\vb{R}} e^{i \vb{k} \cdot \vb{R}} \varphi_{\mu}^{\phantom \dag} (\vb{x} - \vb{R}) \,.
\label{eq:Wannier_def}
\end{equation}
The raising and lowering operators of the Bloch wavefunctions are $\widehat{d}^{\,\, \dag}_{\vb{k}\mu} $ and $ \widehat{d}_{\vb{k}\mu}^{\phantom \dag}$. They are defined by $\widehat{d}^{\,\, \dag}_{\vb{k}\mu} | \Omega \rangle = |\vb{k} \mu \rangle$, $\{ \widehat{d}_{\vb{k} \mu}^{\phantom \dag} , \widehat{d}_{\vb{k}' \mu'}^{\phantom \dag} \} = 0$ and $\{ \widehat{d}_{\vb{k} \mu}^{\phantom \dag} , \widehat{d}_{\vb{k}' \mu'}^{\,\, \dag} \} = \delta_{\vb{k}, \vb{k}'} \delta_{\mu,\mu'}$, where $|\Omega \rangle$ is the vacuum of valence excitations and core holes. A general Hamiltonian describing noninteracting valence electrons is
\begin{equation}
\widehat{\mathcal{H}}_{\text{band}}^{\,} = \sum_{\vb{k} \in \text{BZ}} \sum_{\mu,\mu'} \widehat{d}_{\vb{k}\mu}^{\,\, \dag} \, H_{\mu \mu'}({\vb{k}}) \, \widehat{d}_{\vb{k}\mu'}^{\phantom \dag}  \,, \label{eq:hamiltonian}
\end{equation}
where $H_{\mu \mu'}({\vb{k}})$ are the elements of the matrix $H(\vb{k})$. Let $U(\vb{k})$ be a matrix that diagonalizes $H(\vb{k})$, such that $U^{\, \dag}(\vb{k}) H(\vb{k}) U(\vb{k})$ is a diagonal matrix containing the eigenvalues $\varepsilon_{l}(\vb{k})$, which constitute the dispersing bands. We can then write
\begin{equation}
\widehat{d}_{\vb{k}\mu}^{\phantom \dag} = \sum_{l} U_{\mu l}(\vb{k}) \, \widehat{\psi}_{\vb{k} l}^{\phantom \dag} ,
\label{eq:orbital_to_band}
\end{equation}
where $l$ denotes an energy band index and $\widehat{\psi}_{\vb{k} l}^{\phantom \dag}$ annihilates the corresponding eigenstate. The ground state at zero temperature is obtained by populating all states below the Fermi level:
%\begin{subequations}
\begin{align}
|g \rangle = & \left( \prod\limits_{l, \vb{k} \in \text{BZ}} \Theta (\varepsilon_{\text{F}}^{\,} - \varepsilon_{l}^{\,} (\vb{k})) \, \widehat{\psi}_{\vb{k} l}^{\dag} \right) |\Omega \rangle \,,
\end{align}
%\end{subequations}
%|f \rangle = & \Theta (\varepsilon_{l}^{\,} (\vb{k}_1) - \varepsilon_{\text{F}}^{\,}) \, \Theta (\varepsilon_{\text{F}}^{\,} - \varepsilon_{m}^{\,} (\vb{k}_2)) \, \widehat{d}_{\vb{k}_1 l}^{\dag} \, \widehat{d}_{\vb{k}_2 m}^{\phantom \dag} |g \rangle
with $\Theta$ the Heaviside step function.
The Wannier lowering operators at any lattice site $\vb{R}$ can be expressed in terms of the band operators as
\begin{subequations}
\begin{align}
\widehat{d}_{\vb{R}\mu}^{\phantom \dag} =& \frac{1}{\sqrt{N}} \sum\limits_{\vb{k} \in \text{BZ}} e^{- i \vb{k} \cdot \vb{R}} \, \widehat{d}_{\vb{k}\mu}^{\phantom \dag} \\
=& \frac{1}{\sqrt{N}} \sum\limits_{l, \vb{k} \in \text{BZ}} e^{- i \vb{k} \cdot \vb{R}} \, U_{\mu l}(\vb{k}) \, \widehat{\psi}_{\vb{k} l}^{\phantom \dag}. \label{eq:d_in_terms_of_band}
\end{align}
\end{subequations}

We now assume, as per the fast collision approximation, that the intermediate-state Hamiltonian is well approximated by the band Hamiltonian, along with a core hole inverse lifetime $\Gamma$ in the intermediate-state propagator. Due to this assumption, core electron operators cancel out and the intermediate state propagator becomes simply
\begin{subequations}
\begin{align}
 \widehat{\cal G}(\omega_{\text{in}}^{\,}) =& (E_g + \hbar\omega_{\text{in}}^{\,} + \text{i}\Gamma - \widehat{\cal H}_{\text{band}})^{-1} \,, \\
 =& \sum_{\vb{k}} \sum_l \frac{|\vb{k}, l \rangle \langle \vb{k}, l |}{E_g + \hbar\omega_{\text{in}}^{\,} + \text{i}\Gamma - \varepsilon_{l}^{\,} (\vb{k}) } \,.
\end{align}
\end{subequations}
where $\ket{\vb{k}, l}$ are band eigenstates, and we treat $E_g$ and $\Gamma$ as free parameters to be determined by fitting the x-ray absorption spectrum (see App.~\ref{sec:xas}).

Substituting the expression for $\widehat{d}_{\vb{R}\mu}^{\phantom \dag}$ in Eq.~\eqref{eq:d_in_terms_of_band} in Eq.~\eqref{eq:fundamental}, we obtain the fundamental RIXS scattering amplitude in a band structure
%,\omega_{\vb{k}}
\begin{multline}
F_{\mu \mu'}(\vb{q}, \omega_{\text{in}}^{\,}) = \langle f| \sum_{\vb{k}, \vb{k}' \in \text{BZ}} \sum_{l, l'} \frac{1}{N} \sum_{\vb{R}} e^{-i (\vb{k} + \vb{q} - \vb{k}') \cdot \vb{R}} \\ \times  U_{\mu l}(\vb{k}) \,  U^*_{\mu' l'}(\vb{k}') \, \widehat{\psi}_{\vb{k} l}^{\pd} \, \widehat{\cal G}(\omega_{\text{in}}^{\,}) \, \widehat{\psi}_{\vb{k}' l'}^{\, \dag} | g \rangle \,.
\label{eq:fund_scat_amp}
\end{multline}
Notice that $F$ is independent of the core orbitals at this level of description. The sum over $\vb{R}$ evaluates to $N \delta_{\vb{k}', \vb{k} + \vb{q}}^{\,}$, which enforces $\vb{k}' = \vb{k} + \vb{q}$. When $\varepsilon_{l'}^{\,} (\vb{k} + \vb{q}) > \varepsilon_{\text{F}}^{\,}$, we have that $\widehat{\psi}_{\vb{k} + \vb{q} l'}^{\, \dag} | g \rangle$ is an eigenstate of the band Hamiltonian with energy $E_g^{\,} + \varepsilon_{l'}^{\,} (\vb{k} + \vb{q})$ (otherwise the single particle state is already occupied and this term evaluates to zero). The action of $\widehat{\cal G}(\omega_{\text{in}}^{\,})$ on $\widehat{\psi}_{\vb{k} + \vb{q} l'}^{\, \dag} | g \rangle$ is
\begin{equation}
\widehat{\cal G}(\omega_{\text{in}}^{\,}) \, \widehat{\psi}_{\vb{k} + \vb{q} l'}^{\, \dag} | g \rangle = \frac{\Theta (\varepsilon_{l'}^{\,} (\vb{k} + \vb{q}) - \varepsilon_{\text{F}}^{\,})}{\hbar\omega_{\text{in}}^{\,} - \varepsilon_{l'}^{\,} (\vb{k} + \vb{q}) + \text{i}\Gamma} \, \widehat{\psi}_{\vb{k} + \vb{q} l'}^{\, \dag} | g \rangle.
\end{equation}
Furthermore, the action of $\widehat{\psi}_{\vb{k} l}^{\pd}$ on $\widehat{\psi}_{\vb{k} + \vb{q} l'}^{\, \dag} | g \rangle$ is non-zero only if $\varepsilon_{l}^{\,} (\vb{k}) < \varepsilon_{\text{F}}^{\,}$ (we need this single particle level to be occupied for the term to be non-zero). Using this we obtain
\begin{align}
F_{\mu \mu'}(\vb{q}, \omega_{\text{in}}^{\,}) = \sum_{l, l'} \sum_{\vb{k} \in \text{BZ}} \bigg[ & \langle f | \widehat{\psi}_{\vb{k} l}^{\pd} \widehat{\psi}_{\vb{k} + \vb{q} l'}^{\, \dag} | g \rangle \nonumber \\ 
& \times \Theta (\varepsilon_{l'}^{\,} (\vb{k} + \vb{q}) - \varepsilon_{\text{F}}^{\,}) \nonumber \\ 
& \times \Theta (\varepsilon_{\text{F}}^{\,} - \varepsilon_{l}^{\,} (\vb{k})) \nonumber \\
& \times \frac{U_{\mu l}(\vb{k}) \,  U^*_{\mu' l'}(\vb{k} + \vb{q})}{\hbar\omega_{\text{in}}^{\,} - \varepsilon_{l'}^{\,} (\vb{k} + \vb{q}) + \text{i}\Gamma} \bigg].
\label{eq:final_fund_scat_amp}
\end{align}
The sum over final states $\ket{f}$ can be taken over the eigenstates of $\widehat{\mathcal{H}}_{\mathrm{band}}$. The pair of step functions in the fundamental scattering amplitude given above in Eq.~\eqref{eq:final_fund_scat_amp} ensures that there is a unique $\ket{f}$ that makes the inner product $\langle f | \widehat{\psi}_{\vb{k} l}^{\pd} \widehat{\psi}_{\vb{k} + \vb{q} l'}^{\, \dag} | g \rangle$ non-zero, since the role of the operator pair is to simply create particle-hole excitations across the Fermi level.
%Notice that the set $\{ \vb{k}, \vb{q}, l, l' \}$ uniquely determines a non-trivial $f$ when the step functions are non-zero. 
Thus, the final sum over $|f \rangle$ can be replaced as
\begin{equation}
\sum_f \rightarrow \sum_{l, l'} \sum_{\vb{k} \in \text{BZ}} \Theta (\varepsilon_{l'}^{\,} (\vb{k} + \vb{q}) - \varepsilon_{\text{F}}^{\,}) \, \Theta (\varepsilon_{\text{F}}^{\,} - \varepsilon_{l}^{\,} (\vb{k})).
\end{equation}
This corresponds to summing over final states with one particle-hole excitation in the valence bands. The inner product is then redundant and can be removed.

The final form of the RIXS intensity for systems well-described by band theory is
\begin{widetext}
\begin{align}
I(\vb{q}, \omega_{\text{in}}^{\,}, \Delta \omega, \boldsymbol{\epsilon}_{\text{in}}^{\,}, \boldsymbol{\epsilon}_{\text{out}}^{\,}) = \sum_{l, l'} \sum_{\vb{k} \in \text{BZ}} & \Theta (\varepsilon_{l'}^{\,} (\vb{k} + \vb{q}) - \varepsilon_{\text{F}}^{\,}) \, \Theta (\varepsilon_{\text{F}}^{\,} - \varepsilon_{l}^{\,} (\vb{k})) \nonumber \\
& \times  \left| \sum_{\mu,\nu,\mu'} \langle \mu| \boldsymbol\epsilon_{\text{out}}^{\,} \cdot \widehat{\vb{r}} |\nu \rangle^* \langle \mu'| \boldsymbol\epsilon_{\text{in}}^{\,} \cdot \widehat{\vb{r}} |\nu \rangle \, \frac{U_{\mu l}(\vb{k}) \,  U^*_{\mu' l'}(\vb{k} + \vb{q})}{\hbar\omega_{\text{in}}^{\,} - \varepsilon_{l'}^{\,} (\vb{k} + \vb{q}) + \text{i}\Gamma} \right|^2  \nonumber \\
& \times \frac{\eta}{[\varepsilon_{l}(\vb{k}) - \varepsilon_{l'}(\vb{k}+\vb{q})+ \hbar \Delta \omega]^2 + \eta^2} \,,
\label{eq:rixs_intensity}
\end{align}
\end{widetext}
where we have replaced the Dirac $\delta$-function with a Lorentzian of peak broadening $\eta$ to represent finite experimental resolution.
%Here $\varepsilon_{n}(\vb{k})$ is the dispersion of the $n^{\mathrm{th}}$ band, $\Delta \omega$ is the energy transfer to the material, $U_{\mu l}(\vb{k})$ denotes the band eigenvectors that allow us to transform from orbital basis to band basis (see Eq~\ref{eq:orbital_to_band}), $\Gamma$ is the core-hole broadening and the step functions enforce the condition that the particle sits at an energy level that was respectively previously unoccupied while the hole excitation is at a previously occupied level.

With respect to a local set of coordinate axes, the incoming X-ray polarization $\boldsymbol{\epsilon}$ has components $(\epsilon_x, \epsilon_y, \epsilon_z)$. The outgoing polarization is usually not measured, and hence one sums over either polarizations parallel to and perpendicular to the scattering plane or over left, linear, and right polarizations. We list the polarization matrix elements for the specific case of the $L_3$ edge of a $3d$ transition-metal in Table~\ref{tab:matels}.

\section{Experimental methods}

\subsection{Sample preparation}
Single crystals of FeSi were prepared using a Ga flux method. We mixed the starting materials in a molar ratio of 1:1:15 for Fe:Si:Ga in a glove box filled with argon. This mixture was placed in an alumina crucible and sealed in an evacuated quartz tube. The crucible was heated to 1150$^{\circ}$C and held for 10~h, before cooling to $950^{\circ}$C at 2 K/h, after which the flux was centrifuged. The crystals were washed with diluted hydrochloric acid in order to remove Ga flux from the surface of the samples. This process yielded, samples with large, millimeter-sized flat crystal facets. The samples were checked with lab x-ray diffraction and inelastic x-ray scattering find that the samples have excellent crystallographic quality \cite{Miao2018observation}. The electronic structure of the samples was checked with x-ray absorption spectroscopy during the RIXS experiments, obtaining results that are consistent with the literature \cite{Sirotti1993synchrotron}.

\subsection{RIXS and experimental geometry}
\begin{figure}[ht]
\includegraphics[width=\columnwidth]{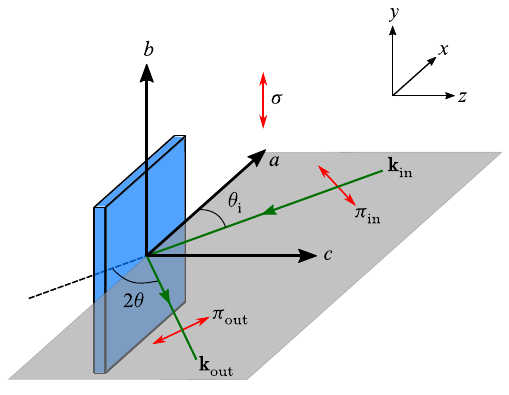}
\caption{Schematic of the RIXS setup. $\vb{k}_{\mathrm{in}}^{\,}$ and $\vb{k}_{\mathrm{out}}^{\,}$ respectively denote the ingoing and outgoing scattering vectors. The components of the ingoing and outgoing photon polarization within the scattering plane are denoted by $\pi_{\mathrm{in}}^{\,}$ and $\pi_{\mathrm{out}}^{\,}$ while the $\sigma$ polarization direction is the same for both. The incident angle $\theta_{\mathrm{i}}^{\,}$ is measured with respect to the sample surface, that is the $a$ direction in the sample coordinates while the $c$ direction is the normal to the sample surface.}
 \label{fig:exp_scheme}
\end{figure}

RIXS measurements were performed at the Soft Inelastic X-Ray (SIX) beamline at the National Syncrotron Light Source-II (NSLS-II).  The energy resolution was 23~meV and data was collected at a temperature of 75 K, a temperature at which FeSi has been observed to behave as a semiconductor with an indirect band gap~\cite{Changdar2020electronic}. The experimental setup is depicted in Fig.~\ref{fig:exp_scheme}. The lab coordinates are denoted by $x,y,z$ while the crystallographic axes are labelled $a,b,c$. We define the incident and outgoing beam angles with respect to the sample coordinate system, wherein the $c$ direction is the normal to the sample surface and the $ac$ plane is the scattering plane. Experimental data are corrected to account for self-absorption effects.

With $\theta_i^{\,}$ denoting the incident angle measured with respect to the sample surface and $2\theta$ denoting the angle between the incident and outgoing beam, we can easily verify the following in the sample frame:
\begin{subequations}
\begin{align}
& \vb{k}_{\mathrm{in}}^{\,} = k_{\mathrm{in}}^{\,} (- \cos \theta_{\mathrm{i}}^{\,}, 0 , - \sin \theta_{\mathrm{i}}^{\,} ) \,,\\
& \vb{k}_{\mathrm{out}}^{\,} = k_{\mathrm{out}}^{\,} \left( -\cos (2 \theta -  \theta_{\mathrm{i}}^{\,}), 0, \sin (2 \theta -  \theta_{\mathrm{i}}^{\,}) \right) \,,\\
& \boldsymbol{\epsilon}_{\pi,\mathrm{in}}^{\,} = (-\sin \theta_{\mathrm{i}}^{\,}, 0, \cos \theta_{\mathrm{i}}^{\,}) \,,\\
& \boldsymbol{\epsilon}_{\pi,\mathrm{out}}^{\,} = (\sin (2\theta - \theta_{\mathrm{i}}^{\,}), 0, \cos (2\theta - \theta_{\mathrm{i}}^{\,})) \,,\\
& \boldsymbol{\epsilon}_{\sigma, \text{in}} = \boldsymbol{\epsilon}_{\sigma, \text{out}} = (0, 1, 0) \,.
\end{align} 
\end{subequations}
Although in reality the ingoing and outgoing photon momentum magnitudes $k_{\mathrm{in}}^{\,}$ and $k_{\mathrm{out}}^{\,}$ are different owing to non-zero energy transfer $\Delta \omega$, the difference is negligible since $\Delta\omega \ll  \omega_{\mathrm{in}}^{\,}$, and hence $k_{\mathrm{in}}^{\,} \approx k_{\mathrm{out}}^{\,} = k$. The momentum transfer \emph{to} the material $\vb{q}$ is then
\begin{equation}
\vb{q} = k \left( \cos (2 \theta -  \theta_{\mathrm{i}}^{\,}) - \cos \theta_{\mathrm{i}}^{\,}, 0, - \sin (2 \theta -  \theta_{\mathrm{i}}^{\,}) - \sin \theta_{\mathrm{i}}^{\,} \right) \,.
\end{equation}
In practice the ingoing and outgoing beam directions are set to specific values, which defines a specific $2\theta$. By rotating the sample about the $y$ axis (or equivalently $b$ axis), we can change $\vb{q}$ by changing $\theta_{\mathrm{i}^{\,}}$.

\section{Density functional theory and tight-binding model}\label{sec:TBM}

\begin{figure}[t]
\includegraphics[width=\columnwidth]{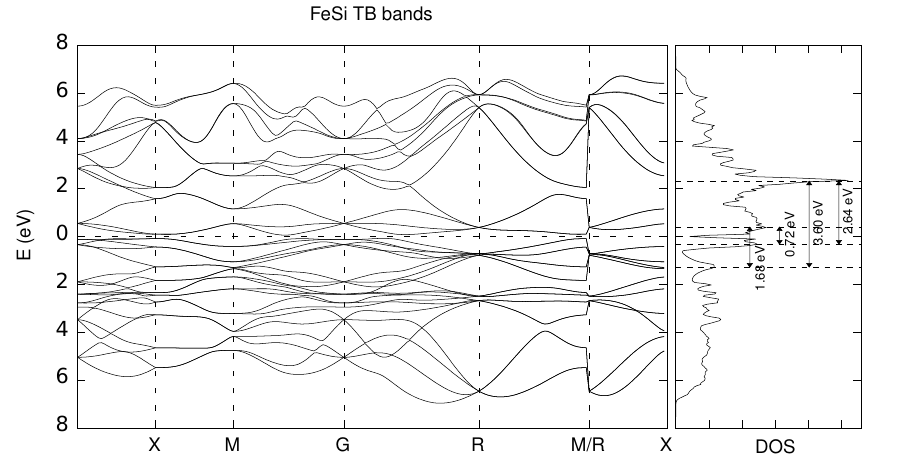}
\caption{Band structure along a high-symmetry path (left) and density of states (DOS) (right) of FeSi as obtained from DFT calculations. Energy differences between most prominent DOS features are indicated with arrows.}
 \label{fig:FeSi_bands}
\end{figure}

The band structure of FeSi was simulated in a similar way to prior studies of FeSi~\cite{Miao2018observation}. We performed first-principles calculations based on DFT~\cite{Hohenberg1964inhomogeneous} within the Perdew-Burke-Ernzerhof exchange-correlation \cite{Perdew1996generalized} implemented in the Vienna ab initio simulation package (VASP)~\cite{Kresse1996effcient}. The plane-wave cutoff energy was 450~eV with a $9\times 9 \times 9$ $k$-mesh in the BZ for self-consistent calculation without considering spin-orbit coupling. Maximally localized Wannier functions~\cite{Marzari1997maximally} were used to obtain the tight-binding model of bulk FeSi with the lattice constants $a = b = c = 4.48$~\AA{}. The calculated band structure and density of states (DOS) are shown in Fig.~\ref{fig:FeSi_bands}.

\section{RIXS spectrum of F\lowercase{e}S\lowercase{i}}\label{sec:results}

\begin{figure}[t]
\includegraphics[width=\columnwidth]{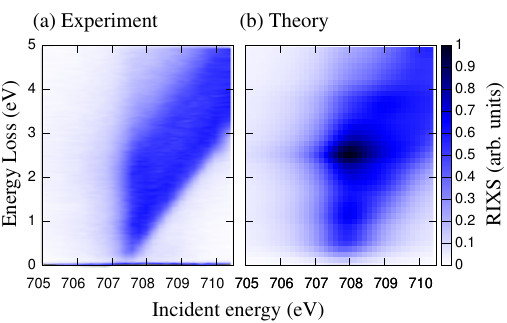}
\caption{Color maps of RIXS intensity with $2\theta = 150^{\circ}$ and $\theta_{\text{i}}^{\,} = 68^{\circ}$ for $\pi$-polarized incident beam as a function of incident photon energy $\hbar\omega_{\mathrm{in}}$ and energy loss $\Delta\omega$ at the $L_3$ edge of Fe in FeSi as obtained (a) in experiment and (b) in the band theory formalism of Sec.~\ref{sec:theory}.}
 \label{fig:thy_RIXS_panel}
\end{figure}

The RIXS intensity at the Fe $L_3$ edge with $\pi$-polarized incident beam is shown in Fig.~\ref{fig:thy_RIXS_panel} in the incident energy-energy loss plane. The absence of prominent sharp inelastic features suggests a particle-hole continuum, consistent with particle-hole excitations in a partially filled band structure.

\begin{figure}[t]
\includegraphics[width=\columnwidth]{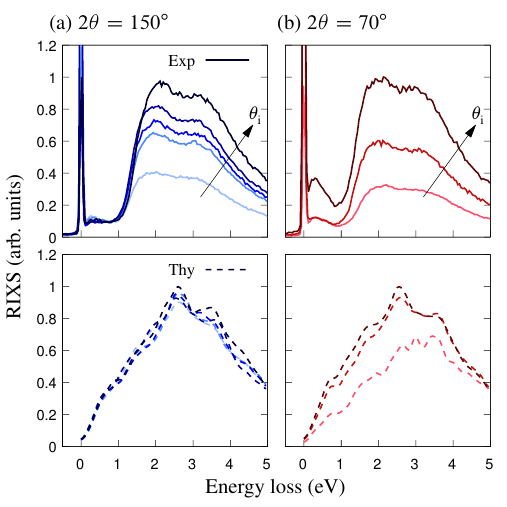}
\caption{Angle-dependent RIXS spectra at $\hbar\omega_{\mathrm{in}} = 708.7$~eV with $\pi$ x-ray polarization for (a) $2\theta = 150^{\circ}$ and (b) $2\theta = 70^{\circ}$ and comparison to simulations within band theory (bottom panels) using Eq.~\eqref{eq:rixs_intensity} with $\Gamma = 0.8$~eV and $\varepsilon_0^{\,} = 707.67$~eV. For scattering angle $2\theta = 150^{\circ}$, the incident angle values (momentum transfer vectors, in units of reciprocal lattice vectors) are $\theta_{\text{i}}^{\,} = 10^{\circ}, 30^{\circ}, 45^{\circ}, 68^{\circ}, 120^{\circ}$ (${\bf q} = (-0.45, 0, 0.21)$, $(-0.35, 0, 0.35)$, $(-0.25, 0, 0.43)$, $(-0.06, 0, 0.49)$, $(0.35, 0, 0.35)$), while for $2\theta = 70^{\circ}$ we have $\theta_{\text{i}}^{\,} = 10^{\circ}, 30^{\circ}, 60^{\circ}$ (${\bf q} = (-0.12,0, 0.27), (-0.03, 0, 0.29), (0.12, 0, 0.27)$).}
 \label{fig:exp_RIXS}
\end{figure}

Angular-dependent RIXS spectra are shown in Fig.~\ref{fig:exp_RIXS} for two values of $2\theta$, which allow us to examine how the spectra change as the momentum and in- and out-going  x-ray polarization are varied. This was performed at an x-ray energy of $\hbar\omega_{\mathrm{in}} = 708.7$~eV with $\pi$ x-ray polarization in order to obtain a strong signal. Spectral weight from inelastic processes lies predominantly in a window of width $\sim 5$~eV. Within that window, all spectra have similar lineshape featuring a sharp onset of intensity at 1~eV, a peak around 2~eV and a shoulder above 3~eV. Overall inelastic intensity increases with increasing $\theta_i$ for both values of $2\theta$. For $2\theta = 150^\circ$, the main peak changes slightly to higher energies with increasing $\theta_{\text{i}}$, from $\sim$2~eV at $\theta_{\text{i}} = 10^\circ$ to $\sim 2.1$~eV at $\theta_{\text{i}} = 120^\circ$, while the shoulder displays no appreciable variation. The RIXS spectra for $2\theta = 70^\circ$ show minimal spectral shape changes with angle. Finally, experimental spectra also contain subdominant features close to the elastic line ($\Delta\omega < 1$~eV)---visible especially in the spectra for $2\theta = 70^\circ$---that, as discussed below, deviate from the band theory predictions.

We use the band theory formulation of Sec.~\ref{sec:theory} and Eq.~\eqref{eq:rixs_intensity} to theoretically model the RIXS process in FeSi~\footnote{The code used for our calculations is provided at \href{https://github.com/anirudhc-git/RIXS_FeSi}{https://github.com/anirudhc-git/RIXS\_FeSi}}. A fit of the absorption spectrum (see App.~\ref{sec:xas}) yields $\Gamma = 0.8$ eV and $E_g = 707.67$~eV, and we choose a peak broadening $\eta = 100$ meV. We use a $48 \times 48 \times 48$ grid of $k$ points in the Brillouin zone for the $32$-band tight binding model detailed above.

The simulated spectra show a structure similar to that observed experimentally, with inelastic weight in a $\sim 5$~eV window containing a peak at $\Delta\omega \sim 2.5$~eV and shoulder at $\Delta\omega \sim 3.5$~eV. As in experiment, overall inelastic intensity increases with increasing $\theta_i$ for both values of $2\theta$, though to a lesser extent. Compared to experiment, features appear at higher energies in simulated spectra. The sharp onset of intensity at 1~eV is not reproduced in calculated spectra; an almost linear increase from 0~eV all the way to the maximum at $\sim 2.5$~eV is seen instead. For both values of $2\theta$, the simulated spectra are essentially the same. For $2\theta = 70^{\circ}$ and $\theta_{\text{i}} = 10^{\circ}$ specifically, the main peak at $\sim 2.5$~eV is absent, leading to theory and experimental spectra that look qualitatively different. Our data include two points in which equivalent momenta of $(\pm 0.35, 0, 0.35)$ and $(\pm 0.12, 0, 0.27)$ are accessed at different angles, providing the opportunity to try to separate momentum- and polarization-dependent effects. We see that since these momentum-equivalent points differ, there polarization-dependent effects are present. The weakness of explicitly momentum-dependent effects may relate to the large number of overlapping electronic bands, as seen in Fig.~\ref{fig:FeSi_bands}, averaging out the momentum dependence in RIXS.

\begin{figure}[t]
\includegraphics[width=\columnwidth]{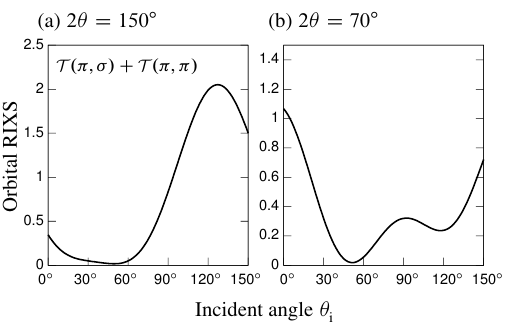}
\caption{The contribution of the dipole matrix elements to the RIXS spectrum of FeSi for $\pi$ ingoing polarisation as given by Eq.~\eqref{eq:rixs_intensity_orbital} after summing over $\sigma$ and $\pi$ polarisations of the outgoing beam.}
 \label{fig:thy_RIXS_polar}
\end{figure}

To investigate what causes the overall increase in RIXS intensity with increasing $\theta_i$, we calculate the atomic scattering tensor~\eqref{eq:geom}. From this we obtain the modulation of the RIXS spectrum~\eqref{eq:rixs_intensity_orbital} coming purely from the orbital content of core and valence states. After summing over outgoing polarizations, we obtain the behavior shown in Fig.~\ref{fig:thy_RIXS_polar}. Comparing to the $\theta_i$ dependence of RIXS spectra in Fig.~\ref{fig:exp_RIXS}, we see that geometric considerations are insufficient to explain the momentum dependence of RIXS intensity in experiment: the momentum dependence of the atomic scattering tensor is different from the one observed, even showing a reverse trend in the case of $2\theta = 70^{\circ}$. Reinstating the band structure fundamental scattering amplitudes in Eq.~\eqref{eq:rixs_intensity} yields the experimentally observed increase of total RIXS intensity with increasing $\theta_i$, albeit only qualitatively, as shown in Fig.~\ref{fig:exp_RIXS}.

\section{Discussion \& conclusion}\label{sec:conclusion}

We have seen that the theoretical formulation of RIXS based on band theory captures the momentum dependence of the total inelastic intensity in experimental Fe $L_3$-edge RIXS spectra of FeSi better than a calculation based on just the atomic multiplet. Band theory also reproduces the right bandwidth for the inelastic part of the spectrum, as well as the most prominent features therein.

On the other hand, non-negligible discrepancies between theory and experiment exist. While the overall momentum dependence of the RIXS spectrum is reproduced by band theory, experimental spectra depend more sensitively on $\theta_{\text{i}}^{\,}$. Spectral features also do not align perfectly between experiment and theory: both peak and shoulder appear at higher energies in calculated spectra. Furthermore, the sharp onset of inelastic intensity at 1~eV is not present in calculations. Finally, a feature around 0.3~eV is present in observed spectra---particularly for $2\theta = 70^{\circ}$---that is absent in our calculations. 

Before we discuss potential reasons for these discrepancies, we emphasize that the level of agreement between band theory and experiment in our work compares favorably to the state-of-the-art of RIXS on approximately noninteracting semimetals~\cite{zhang2012electronic} or topological materials~\cite{jia2022interplay}. Furthermore, all prior work on RIXS of topological materials has been pure theory with no attempt to compare directly to experiment. To our knowledge, our work is the first to attempt a momentum-resolved comparison for this class of materials, in which RIXS is still poorly understood. It is also unique in that it includes the full dipole matrix elements for RIXS of a complex multi-orbital shell in a first-principles approach that includes zero adjustable parameters.

The aforementioned merits of our analysis are a consequence of several approximations made, without which evaluating the RIXS spectrum of FeSi would be intractable. First, the band theory of RIXS ignores electronic correlations. A more detailed theoretical study of the RIXS spectrum would require identifying the precise type, range, and magnitude of interactions present in FeSi. Fully incorporating the effects of interactions in theoretical studies of RIXS is nevertheless a considerable challenge, since we are dealing with a three-dimensional material with 32 relevant orbitals per unit cell. Numerical simulation of the RIXS spectrum based on dynamical mean field theory~\cite{hariki2020lda} may eventually be up to this task. Second, in the fast collision approximation we ignore the effects of a finite core-hole lifetime, which may be appreciable in $3d$ transition metal compounds~\cite{bisogni2014femtosecond}. Future simulations could be improved by incorporating dynamics and interactions with the core hole in the intermediate state.

The main takeaway of our results is that, while RIXS can reveal broad features of the underlying electronic structure in approximately noninteracting materials, a more detailed understanding of mechanisms beyond free-electron physics may be needed to resolve finer features relevant in the study of topological semimetals. In the case of FeSi, the extent to which correlations play a role in the Fe $3d$ shell is not clear. Previous work has revealed signatures of electronic interactions~\cite{Tomczak2012signatures, Wertheim1965unusual} whose importance is temperature-dependent and which may cause substantial band reorganization~\cite{arita2008angle}. Our results are informative in understanding these effects, since they show that the FeSi band structure as obtained by DFT alone cannot fully reproduce the RIXS spectrum.

In conclusion, we have reported RIXS spectra of FeSi at the Fe $L_3$ edge. We observe an excitation continuum without sharp features. Through a band theory formulation of RIXS in the fast collision approximation, we model the RIXS process using the ab initio band structure of FeSi. We obtain reasonable agreement for the spectrum bandwidth, as well as the number and position of main features. Theory also reproduces the angular trend of the overall RIXS intensity, albeit only qualitatively. In contrast, band theory fails to capture the sharp onset of intensity at 1~eV, the subdominant feature at 0.3~eV, and the dispersion of the main peak seen for $2\theta = 150^{\circ}$. The disagreements between theory and experiments are comparable to those seen in recent works on RIXS in semimetals (see, e.g., Ref.~\cite{zhang2012electronic}) and spectroscopy of the bulk in 3D topological materials in general (see, e.g., Refs.~\cite{lv2015observation, zhang2016signatures}). Finally, we emphasize that, in contrast to all previous RIXS studies of semimetals, here we perform a momentum-resolved comparison between experiment and band theory with zero adjustable parameters. Our work thus presents a useful benchmark for future studies that aim to resolve distinctive band structure features in topological materials with RIXS.

\acknowledgments{A.C.~was supported by DOE Grant No. DE-FG02-06ER46316 and EPSRC grant EP/T034351/1. S.K.~acknowledges support from the Minist\`{e}re de l'\'{E}conomie et de l'Innovation du Qu\'{e}bec, a NSERC Discovery grant, and the Canada First Research Excellence Fund. Work at Brookhaven National Laboratory (x-ray scattering and analysis) was supported by the U.S. Department of Energy, Office of Science, Office of Basic Energy Sciences. This research used resources at the SIX beamline of the National Synchrotron Light Source II, a U.S.\ DOE Office of Science User Facility operated for the DOE Office of Science by Brookhaven National Laboratory under Contract No.~DE-SC0012704. We acknowledge National Natural Science Foundation of China (U2032204), the Strategic Priority Research Program of the Chinese Academy of Sciences (XDB33010000) for funding sample synthesis. We thank Yue Cao, Siddhant Das, Claudio Chamon, Michael El-Batanouny, Jungho Kim, and Karl Ludwig for useful discussions.}

\appendix

\section{Self absorption correction}
We performed a self-absorption correction in a similar way to prior Refs.~\cite{Miao2017high,Shen2022role}. In a sample that is thicker than the x-ray absorption length, the effect of self-absorption as 
\begin{equation}
    \frac{I}{I_0} = \frac{\sin \theta_o}{\sin \theta_i + \sin \theta_o},
\end{equation}
where $I$ and $I_0$ are the scattering intensities with and without self-absorption, which depends on incident (scattered) angles $\theta_i$ ($\theta_o=2\theta-\theta_i$). A more sophisticated approach would also account for polarization dependence, but this is expected to be small in a cubic material such as FeSi. 

\section{Polarization matrix elements and the atomic scattering tensor}

The DFT derived tight-binding model used for the calculations presented in the paper involves thirty two basis orbitals per unit cell of the crystal lattice. Due to the assumption of zero spin orbit coupling for the valence bands, this gives rise to thirty-two, two-fold spin-degenerate bands. The orbitals used are the five $3d$ orbitals of each of the four Fe atoms and the three $3p$ orbitals of each of the four silicon atoms within a unit cell, giving a total of 32 orbitals per unit cell.

Since the tight binding model is expressed in terms of $3d$ orbitals whose local axes are perfectly aligned with crystal axis for each of the four Fe atoms in the unit cell, we need to compute the matrix elements of the $2p_{3/2} \rightarrow 3d$ transitions for just one of the atoms. The $2p$ orbitals all have the same radial part of the wavefunction, $\phi_{2p}^{\,} (r)$ and, likewise, the $3d$ orbitals have same radial wavefunction $\phi_{3d}^{\,} (r)$. The radial integral of the various matrix elements in the atomic scattering tensor is simply the radial integral of the product $\phi_{2p}^{\,} (r) \cdot \phi_{3d}^{\,} (r)$ and the radial part of the dipole transition operator. Since this is a common term that just provides an overall multiplicative factor for the RIXS cross section, we ignore it and compute only the azimuthal and polar integrals of the matrix elements. We document the relevant matrix elements of the dipole operator in Table~\ref{tab:matels}, which were verified by comparing to open source RIXS code EDRIXS \cite{Wang2019edrixs}.

\begin{table*}
\begin{center}
\begin{tabular}{c|c|c|c|c}
 & 
$J=\frac{3}{2}, J_z^{\,}=-\frac{3}{2}$ & 
$J=\frac{3}{2}, J_z^{\,}=-\frac{1}{2}$ & 
$J=\frac{3}{2}, J_z^{\,}=\frac{1}{2}$ & 
$J=\frac{3}{2}, J_z^{\,}=\frac{3}{2}$
\\
\hline
$d_{3z^2-r^2 \uparrow}^{\,}$ & 
$\left(0, 0, 0\right)$ & 
$\left(-\frac{1}{6}, \frac{i}{6}, 0\right)$ & 
$\left(0, 0, \frac{2}{3}\right)$ & 
$\left(\frac{1}{2 \sqrt{3}}, \frac{i}{2 \sqrt{3}}, 0\right)$
\\
$d_{3z^2-r^2 \downarrow}^{\,}$ & 
$\left(-\frac{1}{2 \sqrt{3}}, \frac{i}{2 \sqrt{3}}, 0\right)$ & 
$\left(0, 0, \frac{2}{3}\right)$ & 
$\left(\frac{1}{6}, \frac{i}{6}, 0\right)$ & 
$\left(0, 0, 0\right)$
\\
$d_{xz \uparrow}^{\,}$ & 
$\left(0, 0, 0\right)$ & 
$\left(0, 0, \frac{1}{2 \sqrt{3}}\right)$ & 
$\left(\frac{1}{\sqrt{3}}, 0, 0\right)$ & 
$\left(0, 0, -\frac{1}{2}\right)$
\\
$d_{xz \downarrow}^{\,}$ & 
$\left(0, 0, \frac{1}{2}\right)$ & 
$\left(\frac{1}{\sqrt{3}}, 0, 0\right)$ & 
$\left(0, 0, -\frac{1}{2 \sqrt{3}}\right)$ & 
$\left(0, 0, 0\right)$
\\
$d_{yz \uparrow}^{\,}$ & 
$\left(0, 0, 0\right)$ & 
$\left(0, 0, -\frac{i}{2 \sqrt{3}}\right)$ & 
$\left(0, \frac{1}{\sqrt{3}}, 0\right)$ & 
$\left(0, 0, -\frac{i}{2}\right)$
\\
$d_{yz \downarrow}^{\,}$ & 
$\left(0, 0, -\frac{i}{2}\right)$ & 
$\left(0, \frac{1}{\sqrt{3}}, 0\right)$ & 
$\left(0, 0, -\frac{i}{2 \sqrt{3}}\right)$ & 
$\left(0, 0, 0\right)$
\\
$d_{x^2-y^2 \uparrow}^{\,}$ & 
$\left(0, 0, 0\right)$ & 
$\left(\frac{1}{2 \sqrt{3}}, \frac{i}{2 \sqrt{3}}, 0\right)$ & 
$\left(0, 0, 0\right)$ & 
$\left(-\frac{1}{2}, \frac{i}{2}, 0\right)$
\\
$d_{x^2-y^2 \downarrow}^{\,}$ & 
$\left(\frac{1}{2}, \frac{i}{2}, 0\right)$ & 
$\left(0, 0, 0\right)$ & 
$\left(-\frac{1}{2 \sqrt{3}}, \frac{i}{2 \sqrt{3}}, 0\right)$ & 
$\left(0, 0, 0\right)$
\\
$d_{xy \uparrow}^{\,}$ & 
$\left(0, 0, 0\right)$ & 
$\left(-\frac{i}{2 \sqrt{3}}, \frac{1}{2 \sqrt{3}}, 0\right)$ & 
$\left(0, 0, 0\right)$ & 
$\left(-\frac{i}{2}, -\frac{1}{2}, 0\right)$
\\
$d_{xy \downarrow}^{\,}$ & 
$\left(-\frac{i}{2}, \frac{1}{2}, 0\right)$ & 
$\left(0, 0, 0\right)$ & 
$\left(-\frac{i}{2 \sqrt{3}}, -\frac{1}{2 \sqrt{3}}, 0\right)$ & 
$\left(0, 0, 0\right)$
\\
\end{tabular}
\end{center}
\caption{Dipole matrix elements relevant for the $L_3^{\,}$ edge of FeSi. Only the polar and azimuthal integrals are evaluated since the radial integral is the same for all the core-valence pairs, and provides only an overall prefactor to the theoretical RIXS spectrum.}
\label{tab:matels}
\end{table*}

\section{X-ray absorption spectrum and theoretical fit}
\label{sec:xas}

\begin{figure}[ht]
\includegraphics[width=\columnwidth]{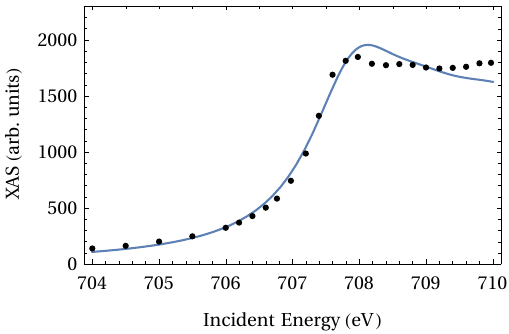}
\caption{Optimal average absolute deviation fit to the $L_3$-edge x-ray absorption spectrum (XAS) of FeSi. The black dots represent the experimental absorption spectrum while the continuous blue line represents the theoretical spectrum calculated using the tight-binding model described in Sec~\ref{sec:TBM}. The fit yields $\Gamma = 0.8$~eV and $E_g = 707.67$~eV.}
 \label{fig:Abs_spect}
\end{figure}

To align the experimental RIXS spectra with theoretical results obtained through ab initio calculations, we need to determine the absolute energy scale $E_g$ of the initial state.
%one has to additionally determine an overall offset $\varepsilon_0^{\,}$, such that the band energies are replaced as $\varepsilon_{l}^{\,} (\vb{k}) \rightarrow \varepsilon_{l}^{\,} (\vb{k}) + \varepsilon_0^{\,}$ and $\varepsilon_{l'}^{\,} (\vb{k} + \vb{q}) \rightarrow \varepsilon_{l'}^{\,} (\vb{k} + \vb{q}) + \varepsilon_0^{\,}$. The reason is that absolute energy scales are not reliably determined within the standard framework of DFT. While this offset cancels out in the step functions and Lorentzian in Eq.~\eqref{eq:rixs_intensity}, it also offsets the ingoing photon energy $\hbar \omega_{\text{in}}^{\,}$ in the intermediate-state propagator.
We determine $E_g$ through a fit of the experimental X-ray absorption intensity with the calculated absorption spectrum
\begin{widetext}
\begin{equation}
I_{\text{abs}}^{\,}(\vb{q}, \omega_{\text{in}}^{\,}, \boldsymbol{\epsilon}_{\text{in}}^{\,}) = \sum_{\boldsymbol{\epsilon}_{\text{out}}^{\,}} \sum_{l, l'} \sum_{\vb{k} \in \text{BZ}} \Theta (\varepsilon_{l'}^{\,} (\vb{k} + \vb{q}) - \varepsilon_{\text{F}}^{\,}) \, \Theta (\varepsilon_{\text{F}}^{\,} - \varepsilon_{l}^{\,} (\vb{k})) \left| \sum_{\mu,\nu,\mu'} \frac{ \langle \mu| \boldsymbol\epsilon_{\text{out}}^{\,} \cdot \widehat{\vb{r}} |\nu \rangle^* \langle \mu'| \boldsymbol\epsilon_{\text{in}}^{\,} \cdot \widehat{\vb{r}} |\nu \rangle \, U_{\mu l}(\vb{k}) \,  U^*_{\mu' l'}(\vb{k} + \vb{q})}{E_g + \hbar\omega_{\text{in}}^{\,} - \varepsilon_{l'}^{\,} (\vb{k} + \vb{q}) + \text{i}\Gamma} \right|^2 \,,
\label{eq:rixs_abs_intensity}
\end{equation}
\end{widetext}
which is obtained by integrating over $\Delta \omega$ the RIXS spectrum in Eq.~\eqref{eq:rixs_intensity}. In addition to $E_g$, we consider the core hole inverse lifetime $\Gamma$ as a tunable parameter in the fit. We sum over outgoing polarizations since the measured spectrum is not polarization resolved. Fig.~\ref{fig:Abs_spect} shows the fit that minimizes the average absolute deviation and yields the values $\Gamma = 0.8$~eV and $E_g = 707.67$~eV.

\bibliography{refs}

\end{document}